\documentclass[prl,amsmath,amssymb,aps,twocolumn,preprintnumbers,showpacs,
floatfix]{revtex4}

\usepackage{color,graphicx} 
\usepackage{float} 
\usepackage{epstopdf} 

\begin{document}

\title{Production of charged heavy quarkonium-like states at the LHC and the  Tevatron }

 \author{Feng-Kun Guo $^a$ } 
 \author{Ulf-G.  Mei\ss ner $^{a,b}$ } 
 \author{Wei Wang $^a$} 

\affiliation{ 
  $^a$ Helmholtz-Institut f\"ur Strahlen- und Kernphysik and Bethe Center for Theoretical Physics,
Universit\"at Bonn, D-53115 Bonn, Germany\\
$^b$ Institute for Advanced Simulation, Institut f\"ur Kernphysik and J\"ulich Center for Hadron Physics,
JARA-FAME and JARA-HPC, Forschungszentrum J\"ulich, D-52425 J\"ulich, Germany}

\begin{abstract}
We  study  prompt hadroproduction  of the charged bottomonium-like states 
$Z_b^\pm (10610)$ and $Z_b^\pm (10650)$, and the charged charmonium-like states 
$Z_c^\pm (3900)$ and $Z_c^\pm (4020)$, at the Tevatron and the LHC, provided that 
these   states are   $S$-wave  hadronic molecules.  Using two Monte Carlo event 
generators,  Herwig and Pythia, to simulate the production of  heavy meson pairs,  
we derive an order-of-magnitude estimate of the production rates for these four 
particles.  Our estimates
yield a cross section at the   nb level for the $Z_b(10610)$ and $Z_b(10650)$. 
The results for the $Z_c(3900)$ and $Z_c(4020)$ are larger by a factor of 
$20-30$.  These cross sections are large enough to be observed, and 
measurements at hadron colliders in the future will supplement the study using 
electron-positron collisions,  and therefore allow to explore  
the mysterious nature of  these exotic states.  

\end{abstract}

\pacs{13.85.Ni;14.40.Rt }
\maketitle

{\it Introduction} -- In the past decade, a number of  $X,Y,Z$  states have 
been discovered at various experimental facilities,  many of which may be exotics  
beyond  the quark 
model~\cite{Brambilla:2010cs}. Among these  states,  charged charmonium-like 
and bottomonium-like ones are of particular interest due to their genuine
four-quark nature.

In 2011, the Belle Collaboration  reported a measurement of the $\pi^\pm 
\Upsilon(nS)(n=1,2,3)$ and $\pi^\pm h_b(mP) (m=1,2)$ invariant mass spectra in 
the processes $e^+e^-\to \Upsilon (5S)\to  \Upsilon(nS) \pi^+\pi^-$ and $ 
e^+e^-\to \Upsilon (5S)\to h_b(mP)\pi^+\pi^-$~\cite{Belle:2011aa}, in which two unexpected
charged bottomonium-like structures $Z^\pm_b(10610)$ and $Z^\pm_b(10650)$ were 
discovered. Their masses from a fit using a Breit-Wigner type 
parameterization are very close to the $B\bar B^*$ and $B^*\bar B^*$ 
thresholds, respectively. The angular distribution analysis indicates that 
their quantum numbers   are $I^{G}(J^P)=1^+(1^+)$. 
Recently, the BESIII and Belle Collaborations have  studied the process 
$e^+e^-\to  \pi^+\pi^- J/\psi$ at the center-of-mass energy around $4.26$ GeV, and 
found  a charged charmonium-like state $Z_c^\pm(3900)$, with the central value 
of the measured mass being about 20 MeV above the $D\bar D^*$ threshold~\cite{Ablikim:2013mio,Liu:2013dau}. The 
observation was  confirmed 
later on by an analysis based on the CLEO data at the 
energy of 4.17 GeV~\cite{Xiao:2013iha}. Just above 
the $D^*\bar D^*$ threshold, there is evidence for another charged structure 
from the BESIII data in the $h_c\pi^\pm$ and $D^*\bar D^*$ mass 
distributions~\cite{Zc_4020}. The masses and widths of these four states are 
collected in Tab.~\ref{tab:dataZbZc}. Apparently all of them defy a standard 
bottomonium/charmonium assignment, as they should consist 
of at least four  quarks, e.g. $bu\bar{b} \bar{d}$ or $cu\bar{c} \bar{d}$ for 
the positively charged states.

Due to the fact that the masses of the discovered states lie in the vicinity of  
meson-meson thresholds, it has been suggested  that they  are $S$-wave 
molecular states of heavy meson anti-heavy meson pairs with nearby 
thresholds, i.e. the $B^{(*)}\bar{B}^*$ and the
$D^{(*)}\bar{D}^*$ (see  
Refs.~\cite{Bondar:2011ev,Sun:2011uh,Cleven:2011gp,Cleven:2013sq,Wang:2013cya,
Guo:2013sya, Wilbring:2013cha} and many references therein). 
The fact that  these four particles have a typical hadronic  total width of a 
few tens of MeV has also triggered another four-quark interpretation,  in which 
they are viewed as genuine four-quark states (tetraquarks)  
compactly bound by gluons (see for example 
Refs.~\cite{Ali:2011ug,Ke:2012gm,Faccini:2013lda,Braaten:2013boa} and references therein).

\begin{table}[t]
\caption{Experimental data for the mass and width (in units of MeV) of the 
$Z_b(10610)$, $Z_b(10650)$~\cite{Adachi:2012cx}, $Z_c(3900)$ (average of  
results from Refs.~\cite{Ablikim:2013mio,Liu:2013dau,Xiao:2013iha}), and 
$Z_c(4020)$ (average of the  results from  Ref.~\cite{Zc_4020}). }
\label{tab:dataZbZc}
\vspace{-8pt}
\begin{center}
\begin{ruledtabular}
\begin{tabular}{lcccc}
 & $Z_b(10610)$ & $Z_b(10650)$ &$Z_c(3900)$& $Z_c(4020)$\\ \hline
  mass & $10607.2\pm 2.0$& $10652.2\pm 1.5$ & $3891.5\pm 3.5$ &$4023.0\pm 2.3$  \\
  width &$18.4\pm 2.4$& $11.5\pm 2.2$  &$39.2\pm 10.5$ & $9.7\pm 3.2$\\
\end{tabular}
\end{ruledtabular}
\end{center}
\vspace{-20pt}
\end{table} 

Hidden charm hadrons  can be produced at $e^+e^-$ machines not only directly, but also in $B$ decays and radiative return studies.
The study of bottom hadrons highly relies on direct measurements, most of which 
have been tuned to the center-of-mass energy of the masses of the $\Upsilon(4S)$ and 
the $\Upsilon(5S)$. The data set for the latter has much smaller statistics.  
The situation will be greatly improved in the future as the Super KEKB factory   will provide a large amount of data. 

On the other hand, since bottom and charm quarks will be abundantly produced at hadron colliders 
with very high  luminosity like at the LHC and the Tevatron, hadroproduction 
processes  can provide complementary  and independent checks 
on these mysterious states, and help in understanding their underlying nature. 
Furthermore, the study of production rates of  heavy  hadronic molecules at 
hadron colliders can provide much information  to 
understand the interplay of perturbative and nonperturbative QCD effects. 
For instance,  the $X(3872)$,  an ideal example of a loosely-bound hadronic molecule,  can be copiously produced in high energy collisions at the 
Tevatron~\cite{Abazov:2004kp,Aaltonen:2009vj} and 
LHC~\cite{Chatrchyan:2013cld,Aaij:2013zoa}, for which different  theoretical 
predictions were made in 
Refs.~\cite{Bignamini:2009sk,Artoisenet:2009wk,Esposito:2013ada,
Artoisenet:2010uu}.

In this paper, we will  derive for the first time an estimate of the  prompt 
production rates of these four charged particles, the $Z_b^\pm(10610)$,  
$Z_b^\pm(10650)$, $Z_c^\pm(3900)$ and $Z_c^\pm(4020)$,  in proton-(anti)proton 
collisions at the LHC and Tevatron, assuming them to be hadronic molecules.  
``Prompt'' means that these states are not produced from the decays of 
particles of higher masses. 
The production of the $Z_b^\pm(10610)$ and $Z_b^\pm(10650)$ from the decays of 
the $\Upsilon(5S)$ was discussed in Ref.~\cite{Ali:2013xba}.  For the sake of 
simplicity, the four states will be abbreviated as $Z_Q$ with the mesonic 
constituents $H\bar H$, where the heavy quark is denoted as $Q$ and $H(\bar H)$ 
represents the relevant heavy (anti-)meson.

{\it Hadroproduction of hadronic molecules} --
The production rates of $S$-wave hadronic molecules can be quantitatively 
calculated  under  a few assumptions:
\begin{itemize}
\item The production rates of hadronic molecules satisfy the factorization 
ansatz, which will separate the formation of a molecule at  long distance from 
the short-distance production of its constituents. 
\item
A hadronic molecule can be formed only if its constituents are produced with a
relative momentum less than some critical value. 
\end{itemize}

These assumptions  lead to the simplification  that the production 
rates for $Z_Q$ can be expressed in terms of the cross sections for
the inclusive production of $H\bar H$ (and charge conjugates). As a 
phenomenological and successful  tool that has been utilized  in many other 
processes, Monte Carlo (MC) event generators are able to simulate the 
hadronization of partons produced in the QCD processes, and therefore provide an 
 estimate of  the $pp/\bar p\to H\bar H$ inclusive cross sections. 

At the energy region around the pole of a narrow resonance, the resonance will 
provide a dramatic energy dependence of the relevant differential cross 
sections. Around the $H\bar H$ threshold, one may neglect the impact of the 
inelastic channels and apply  Watson's theorem~\cite{Watson:1952ji} to relate 
the $H\bar H$ final-state-interaction (FSI) in the production to the $H\bar H$ 
scattering amplitude. Thus, when the relative momentum of the two constituents is small, 
the momentum dependence of the amplitude for the production of molecules arises 
from the $S$-wave scattering amplitude~\cite{Artoisenet:2009wk}. For a narrow 
resonance, it may be effectively  described by a Breit-Wigner parameterization
\begin{eqnarray}
 f(E)= \frac{1}{8\pi m_{Z_Q}}\frac{|g|^2}{ E^2-m^2 +i m_{Z_Q}\Gamma_{Z_Q}(k)}, 
\end{eqnarray}
where $g$ is the $Z_Q H\bar H$ coupling constant, and $E$ is the energy of 
the two constituents. Assuming that the total width is saturated by the 
decay $Z_Q\to H\bar H$, we have
\begin{eqnarray}
 \Gamma_{Z_Q}(k)= \frac{|\vec k|} {8\pi m_{Z_Q}^2 } |g|^2, 
\end{eqnarray}
with $k$ the center-of-mass momentum of the $H$.

The factorization ansatz allows for the separation of the long-distance and 
short-distance contributions in the amplitudes for the production of the molecules. 
The latter is the same for the processes $pp/\bar p\to H\bar H$ and $pp/\bar 
p\to Z_Q$, while the long-distance factor can be deduced from the scattering 
amplitude given above. Following Ref.~\cite{Artoisenet:2009wk}, we assume that 
beyond the resonance region, the production rate is given by the MC event 
generators. This is achieved by matching the production amplitude including the 
FSI effect to the MC one at an energy $E\approx E_\Lambda \equiv m_{Z_Q} + 
\Gamma_{Z_Q}$ as follows
\begin{eqnarray}
\sigma (Z_Q) \approx K_{H\bar H} \int_0^{\Lambda} d k 
\frac{d\sigma_{H\bar H}^{\rm MC} (k)}{ dk } 
\frac{|f(E)|^2}{|f(E=E_\Lambda)|^2},  \label{eq:dsigmadk}
\end{eqnarray} 
where $d\sigma_{H\bar H}^{\rm MC} (k)/ dk$ is the differential cross section of 
the $H\bar H$ inclusive production, and $K_{H \bar H}\sim {\cal O}(1)$ is a 
normalization  factor that  is introduced to compensate the overall difference 
between the MC simulation and the experimental data. The cut-off $\Lambda$ in 
the momentum integration is the center-of-mass momentum evaluated at the 
energy value $E=E_\Lambda$. In case that the considered molecule is a bound 
state, the above formula can be reduced to the  form  derived for the process $pp\to 
X(3872)$ in Ref.~\cite{Artoisenet:2009wk}.

To form a molecular state, the constituents  must move nearly collinear and 
have a small relative momentum.  Such configurations can be realized in an
inclusive  QCD  process which contains  a  $\bar QQ$ pair with approximately the 
same relative momentum in the final state. However, it is necessary to stress 
that since the  $Z_Q^\pm$ consist of at least four quarks, the final state  
must contain at least six quarks: $\bar QQ \bar uu \bar dd$ with  $\bar QQ\bar u 
d$ (or charge conjugates) moving coherently.   At the hadron level, the 
additional light quarks form one or more pions. These multiquark final states 
can be produced by the soft parton shower radiations  in the $2\to 3$ QCD  
events. The dominant partonic process is $gg\to \bar QQ g$, as the gluon density 
at  the LHC and Tevatron energy is much  larger than  those for quarks. 
Nevertheless, other contributing processes are also included in this analysis.

{\it Results} --
We use  Madgraph~\cite{Alwall:2011uj} to generate the  $2\to 3$  partonic events 
having $\bar bb$ (or $\bar cc$) in the final states, and then pass them to the 
MC event generators to  hadronize.   At the parton level, to increase 
the efficiency, we apply a cut  $p_T>2$~GeV for heavy quarks and light jets,  
$m_{b\bar b}< 11$~GeV (corresponding to $k_{B\bar B^*}=1.5$~GeV and $k_{B^*\bar 
B^*}=1.4$~GeV at the hadron level), $m_{c\bar c}< 5$~GeV  (corresponding to 
$k_{D\bar D^*}=1.6$~GeV and $k_{D^*\bar D^*}=1.5$~GeV)
and $\Delta R(b, \bar b)<1$ ($\Delta R(c, \bar c)<1$)  with  $\Delta 
R=\sqrt{\Delta \eta^2 +\Delta \phi^2}$, where $\Delta \phi$ is the azimuthal 
angle difference and $\Delta \eta$ is the pseudo-rapidity difference between 
the $Q$ and the $\bar Q$.  Since additional light quark pairs are produced apart from 
the $\bar QQ$ (corresponding to  pion radiation at the hadron level), a too 
stringent cut on $m_{Q\bar Q}$  may underestimate the production 
rates of $H\bar 
H$. On the other hand, this cut cannot be too large, otherwise the efficiency 
of the numerical calculation will be highly reduced.  The above  choice is a 
compromise and will not affect the $H\bar H$ production rates in the region of 
interest for the study of molecular states.  We choose  
Herwig~\cite{Bahr:2008pv} and Pythia~\cite{Sjostrand:2007gs} as the  
hadronization generators, whose output   are further analyzed 
using the Rivet library~\cite{Buckley:2010ar} and at this step we select 
the heavy meson pair with the smallest invariant mass.

\begin{figure*}[!t] 
\includegraphics[width=0.45\textwidth]{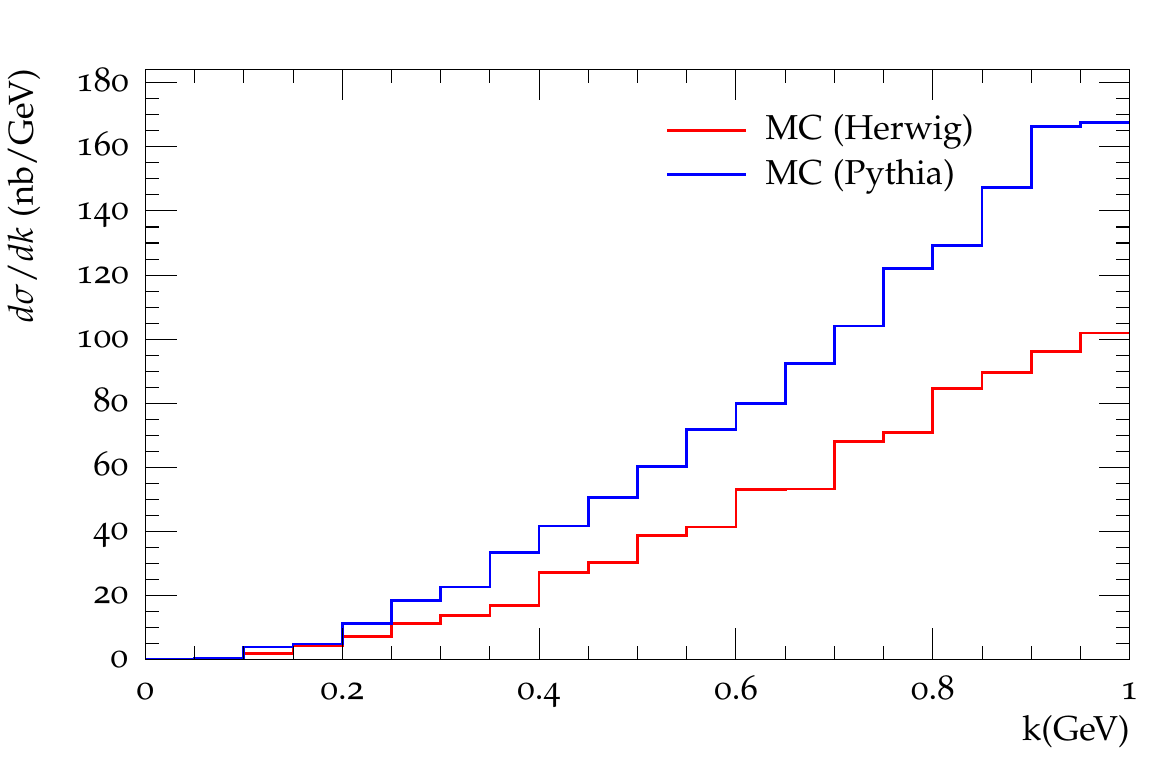}
\includegraphics[width=0.45\textwidth]{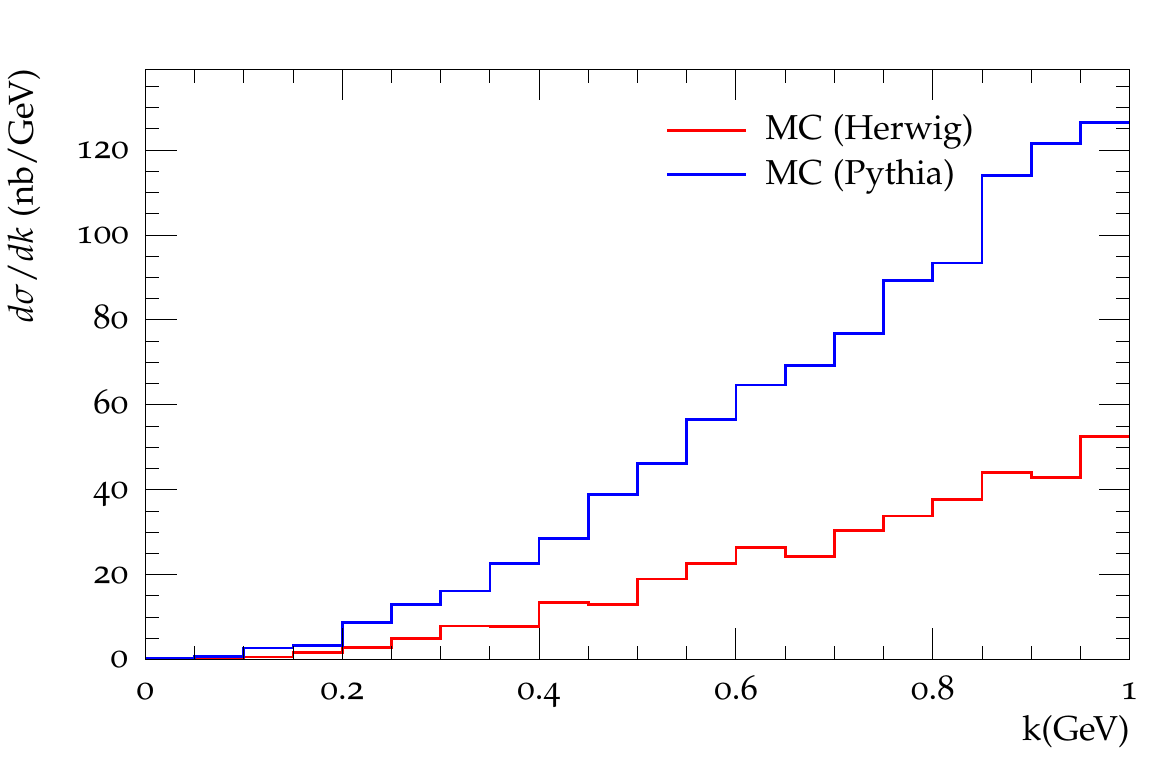} 
\includegraphics[width=0.45\textwidth]{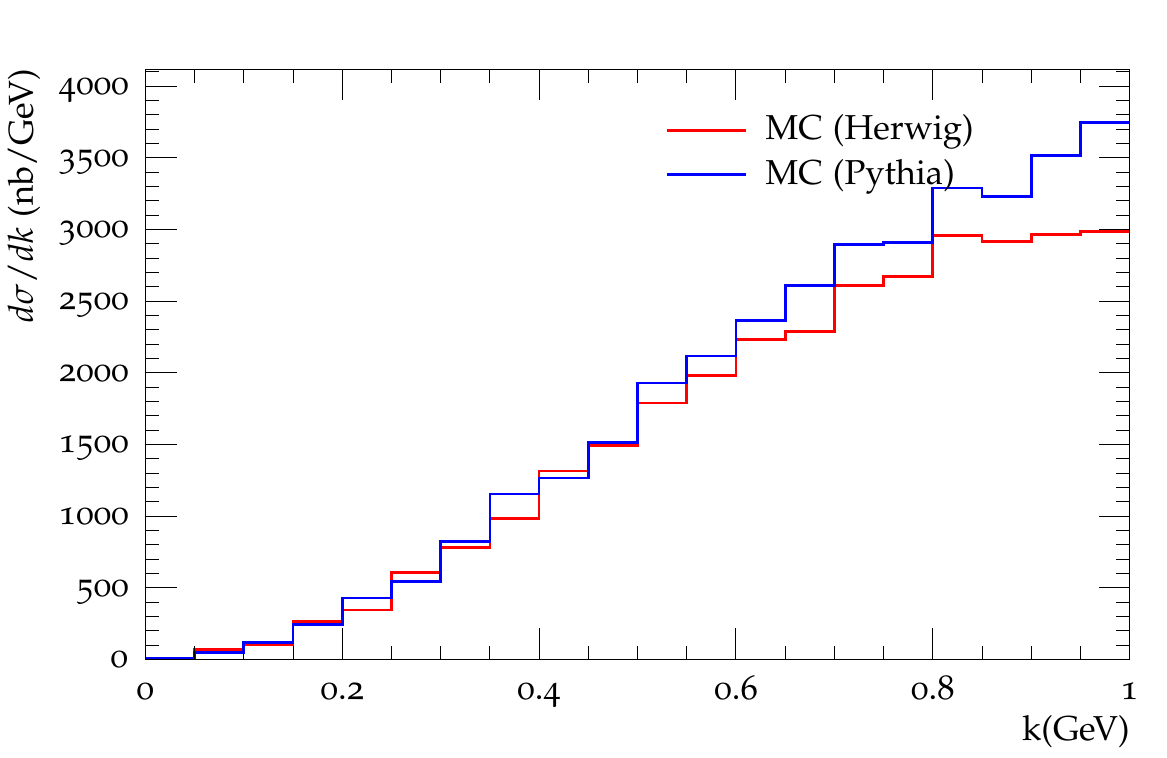}
\includegraphics[width=0.45\textwidth]{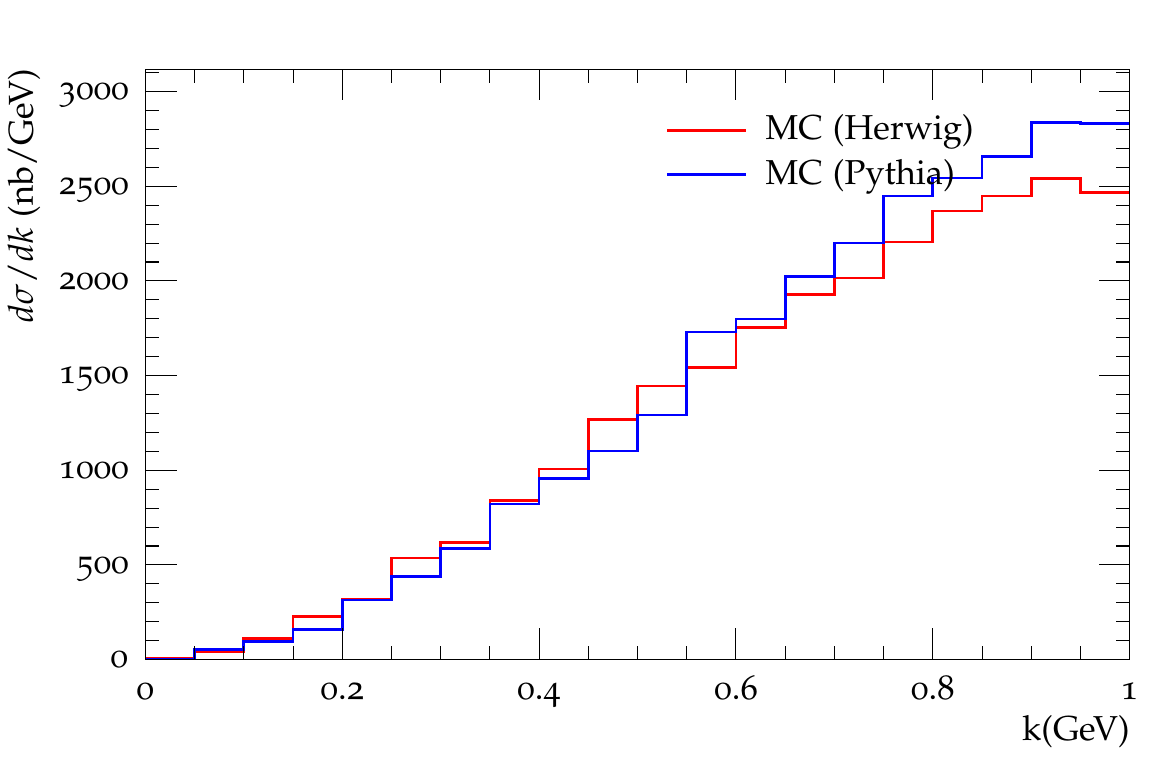} 
\caption{\label{fig:mass_LHC} 
Differential cross sections ${d\sigma}/{ dk}$  (in units of nb/GeV) 
for the inclusive processes $pp\to B^+\bar B^{*0}, pp\to B^{*+}\bar B^{*0}$ 
(the upper two panels) and $pp\to   D^+\bar D^{*0}, pp\to D^{*+}\bar D^{*0}$ 
(the lower two panels) at the LHC with $\sqrt s=8$ TeV.  The kinematic cuts 
used are   $|y|<2.5$ and $p_T>5$ GeV, which lie in the phase-space regions of the 
ATLAS and CMS detectors. }
\end{figure*}

Based on the data sample collected  at 7~TeV, the inclusive differential  cross 
sections for $pp\to B\bar B$ have been measured by the CMS 
Collaboration~\cite{Khachatryan:2011wq}.  It is found that  the cross sections 
are substantial at small values of  $\Delta \phi$ and $\Delta R$, in the same 
kinematics region considered in this work. We wish to point out that the 
simulation in Pythia is in qualitative  consistency  with the experimental data 
for most observables for the $B\bar B$ angular 
correlation~\cite{Khachatryan:2011wq}.
The overall  agreement may, at least qualitatively,  verify $K_{B\bar B^*}\sim  K_{B^*\bar 
B^*}\sim 1$ for the current analysis,  as our main concern is to estimate the 
production cross sections at the order-of-magnitude level.

We have generated $ 10^7$ partonic events, based on which we  show the 
differential distribution $d\sigma/dk$ (in units of nb/GeV)  for the inclusive 
processes $pp\to B\bar B^*$ (upper left panel) and  $pp\to B^*\bar B^*$ (upper 
right panel) at the LHC collider with $\sqrt s= 8$ TeV in 
Fig~.\ref{fig:mass_LHC}. The lower panels correspond to the distributions for 
the $pp\to D\bar D^*$ (left) and the $pp\to D^*\bar D^*$ (right).  The 
kinematic cuts are chosen  in accordance with the default choice of  the 
CMS/ATLAS   measurements: $p_T> 5$ GeV and $|y|<2.5$.

\begin{table}[t]
\caption{Integrated normalized cross sections (in units of nb)
for the reactions $pp/\bar p\to Z_b(10610),Z_b(10650),Z_c(3900)$, 
and $Z_c(4020)$ at the LHC and the Tevatron.
Results are obtained using Herwig (Pythia).
The rapidity range $|y| <2.5$ 
has been assumed for the LHC experiments 
(ATLAS and CMS) at 7, 8 and 14~TeV, respectively, 
for the Tevatron experiments 
(CDF and D0) at 1.96 TeV, we use $|y| <0.6$;
the rapidity range $2.0<y<4.5$ is used for  LHCb.}
\label{tab:integratedCrossSection}
\begin{ruledtabular}
\begin{tabular}{lccccc}
& $Z_b(10610)$ & $Z_b(10650)$ & $Z_c(3900)$ & $Z_c(4020)$ \\\hline
Tevatron
 &0.26(0.47)
 &0.06(0.17)
 &11(13)
 &1.7(2.0)
\\
LHC 7
 &4.8(8.0)
 &1.2(3.0)
 &187(211)
 &29(31)
\\
LHCb 7
 &0.76(1.3)
 &0.18(0.47)
 &33(39)
 &5.5(5.8)
\\
LHC 8
 &5.9(9.5)
 &1.4(3.5)
 &220(240)
 &34(36)
\\
LHCb 8
 &0.9(1.4)
 &0.22(0.56)
 &40(48)
 &6.3(6.9)
\\
LHC 14
 &11(17)
 &2.6(6.5)
 &382(423)
 &61(63)
\\
LHCb 14
 &1.9(3.0)
 &0.52(1.2)
 &84(88)
 &14(14)
\\
\end{tabular} 
\end{ruledtabular}
\end{table}


Using Eq.~\eqref{eq:dsigmadk}, we  show the integrated cross sections for the 
production of molecules in Tab.~\ref{tab:integratedCrossSection}.  We also 
modify  the rapidity range to  $2 .0<y<4.5$ in order to match the
specifics
of the LHCb detector. For the  Tevatron experiments 
(CDF and D0) at 1.96 TeV, we use $|y| <0.6$, which is the 
choice for the measurement of the $p\bar p\to X(3872)$.

The charged $Z_b(10610)$ and $Z_b(10650)$ have a large decay branching
fraction into $\Upsilon(nS)\pi^\pm$,  and thus  can be reconstructed in the 
$(\mu^+\mu^-)\pi^\pm$ final states.  
Using the decay branching fractions~\cite{Adachi:2012cx,Beringer:1900zz},  
we find that the  cross sections for the $pp\to Z_b(10610)^\pm 
\to \Upsilon(2S)\pi^\pm \to  \mu^+\mu^-\pi^\pm$  and the $pp\to Z_b(10610)^\pm 
\to \Upsilon(3S)\pi^\pm \to  \mu^+\mu^-\pi^\pm$ can reach ${\cal 
O}(10~\text{pb})$ at the LHC. To estimate the number of events, we take the 
integrated luminosity of  22 fb$^{-1}$ collected by ATLAS in 
2012~\cite{ATLAS:luminosity} (similar for CMS~\cite{CMS:luminosity}), and this  
yields  ${\cal O}(10^5)$ events for these processes, respectively. At the Tevatron 
and LHCb, the number of events is smaller by about one order-of-magnitude. The 
$Z_c(3900)$ and $Z_c(4020)$ will be more copiously produced, since the cross 
sections  for the $pp\to Z_c(3900)$ and $pp\to Z_c(4020)$ are larger by a factor 
of $20-30$ than those for the $Z_b(10610)$ and $Z_b(10650)$. The differences 
between the results derived from the two MC events reflect the hadronic 
uncertainties.

The production rates for  $Z_b$ states at the LHC predicted in this work, of 
the order a few nb,  are much larger than those non-prompt rates from the 
decays of the $\Upsilon(5S)$ or an exotic candidate 
$Y_b(10890)$~\cite{Ali:2013xba,Ali:2011qi}.  The cross section for the latter 
process $pp\to \Upsilon(5S)\to Z_b\pi$ is predicted to be at the  pb 
level~\cite{Ali:2013xba}, and this finding can be examined in future 
experiments.

{\it Summary} -- In conclusion, we  have studied  the prompt production  of 
charged bottomonium-like  and charmonium-like states discovered in $e^+e^-$ 
annihilation experiments, the $Z_b(10610)$, $Z_b(10610)$, $Z_c(3900)$ and 
$Z_c(4020)$ states,  at the Tevatron and the LHC.  These four particles are 
candidates of hadronic molecules formed of a pair of heavy mesons. We have used 
two Monte Carlo event generators,  Herwig and Pythia, to simulate the 
hadronization, based on which  we found that the inclusive cross sections for 
the $pp/\bar p\to Z_b(10610)/Z_b(10650)$ are of the order of nb, and  results for the
$Z_c(3900)$ and $Z_c(4020)$ are larger by a factor of $20-30$. Taking into 
account  the current integrated luminosities at the LHC, the number of events has be estimated.  
Measurements at hadron colliders will supplement the studies performed at 
electron-positron colliders, and thus allow to  explore the nature of  these four exotic states.

{\it Acknowledgments} -- W.W.  is grateful  to Ahmed Ali, Pierre Artoisenet  
and Christian Hambrock  for enlightening  discussions, to  Simon Pl\"atzer for 
constant help in using the Herwig++,  and to Sebastian K\"onig for arranging the 
computing resources. This work is supported in part by the DFG and the NSFC 
through funds provided to the Sino-German CRC 110 ``Symmetries and the Emergence 
of Structure in QCD'', and by the NSFC (Grant No. 11165005).  More detailed results  
 are available at:\\
\url{http://www.itkp.uni-bonn.de/~weiwang/hadronLHC.shtml}.

\end{document}